\documentclass[hidelinks,12pt]{article}
\usepackage{setspace,amsmath,amssymb,amsthm,mathrsfs,sectsty,hyperref,xcolor,graphicx,geometry,braket,authblk,rotating,lscape}
\usepackage[square,sort,comma,numbers]{natbib}
\hypersetup{
    colorlinks,
    linkcolor={red!50!black},
    citecolor={red!50!black},
    urlcolor={blue!50!green}
}
\allsectionsfont{\centering}
\begin{document}

\title{Contagion processes on urban bus networks in Indian cities}

\author[1]{\small Atanu Chatterjee\thanks{atanu@wpi.edu}}
\author[2]{Gitakrishnan Ramadurai}
\author[3]{Krishna Jagannathan}
\affil[1]{Department of Physics, Worcester Polytechnic Institute, 100 Institute Road, Worcester, MA, 01609, USA}
\affil[2]{Department of Civil Engineering, Indian Institute of Technology Madras, Chennai 600036, India}
\affil[3]{Department of Electrical Engineering, Indian Institute of Technology Madras, Chennai 600036, India }

\date{}

\maketitle

\begin{abstract} 
\noindent Bus transportation is considered as one of the most convenient and cheapest modes of public transportation in Indian cities. Due to their cost-effectiveness and wide reachability, they help a significant portion of the human population in cities to reach their destinations every day. Although from a transportation point of view they have numerous advantages over other modes of public transportation, they also pose a serious threat of contagious diseases spreading throughout the city. The presence of numerous local spatial constraints makes the process and extent of epidemic spreading extremely difficult to predict. Also, majority of the studies have focused on the contagion processes on scale-free network topologies whereas, spatially-constrained real-world networks such as bus networks exhibit a wide-spectrum of network topology. Therefore, we aim in this study to understand this complex dynamical process of epidemic outbreak and information diffusion on the bus networks for six different Indian cities using SI and SIR models. We identify epidemic thresholds for these networks which help us in controlling outbreaks by developing node-based immunization techniques.
\end{abstract}
%
%
\section{Introduction}
The earliest accounts of mathematical modelling to capture the spread of diseases dates back to as early as the $17^{th}$ century. Bernoulli used mathematical equations to defend his stand on vaccination against the outbreak of smallpox. Works following Bernoulli\rq{s} earliest formulation of epidemic modelling helped in understanding germ theory in detail. However, this was not until the works of McKendrick and Kermack, which first proposed a deterministic model that predicted epidemic outbreaks very similar to the ones that were recorded during those times~\cite{kermack1927contribution}. Since then our understanding of the mathematical models in epidemology has evolved over the years, the accounts of which can be found in the extensive works of Anderson and May~\cite{anderson1992infectious}. All the above formulations focused on modelling epidemics over a set of population in which uniform ties between agents were assumed \emph{a priori}. Also, the nature of the ties in the above models did not play any significant role. Contrary to this, the enormous body of work during the last decade in the field of network science has asserted the fact that ties, their strength and types in a system or a population play a significant role~\cite{albert2002statistical, granovetter1973strength, watts1998collective}. Thus, a system or a population is not considered to be a set of individuals, rather it is considered to be a complex network of interacting individuals, where each individual or an entity is considered to be a node, and the links between them define the type of relationship one shares with the other. Interestingly, a network model is not exhaustive to the study of population, rather it is a universal framework which can be used to understand numerous complex systems in general. Also, over the years the term \emph{epidemic modelling} has evolved into a common metaphor for a wide array of dynamical processes on these networks. Various complex phenomena such as percolation, the spreading of blackouts to the spreading of memes, ideas and opinions in a social network can be modelled under the common framework of epidemic modelling~\cite{vespignani2009predicting, bale2014modeling, schuster2015ebola}. 

Transportation networks play a vital role in the spread of epidemics due to their widespread outreach across cities, countries and continents~\cite{pastor2001epidemic}. \lq\lq{S}hould people be worried about getting Ebola on the subway$?$\rq\rq{,} was one of the numerous similar headlines that made the front pages of the newspapers around the world during the 2014 Ebola scare~\cite{paper}. In this particular incident however, nobody was infected because the subject did not show symptoms of Ebola while using public transportation. Therefore not only airline networks, that can transmit pathogens across continents, even modes of public transport operating within cities such as, buses and subways pose a serious threat as well as a source of panic during desperate times. Although, epidemic spreading in airline networks have been studied extensively, similar studies on bus networks are relatively rare~\cite{albert2002statistical, pastor2001epidemic}. Epidemological models have been simulated on bus network datasets; however, the results were only used to validate the numerical models. Also, a recent study on city-wide Integrated Travel Networks (ITN), have found both the traveling speed and frequency to be important factors in epidemic spreading~\cite{ruan2015integrated}. Thus, the effect of network structure and constraints in epidemic spreading are yet to be studied in these networks. In this paper we exploit the structural aspect of these networks to understand their role in epidemic spreading and diffusion process in detail. 

\section{Methodology}
In this paper, we simulate the epidemic model on the bus networks of six major Indian cities, namely Ahmedabad (ABN), Chennai (CBN), Delhi (DBN), Hyderabad (HBN), Kolkata (KBN) and Mumbai (MBN)~\cite{chatterjee2014scaling, chatterjee, chatterjee2015studies}. The route data for all these bus networks are obtained from their respective state government websites. A bus network in $L$-space can be considered a graph, $G = (N, L)$, where $N$ denotes the set of nodes and $L$, the set of links. The topological structure of the graph is generated by considering every stop as a node, and the routes connecting the stops form the set of links. Thus, a bus network $G$ can be represented as a $N\times N$ adjacency matrix, $A_{ij}$ whose elements $a_{ij} = 1$, if $i$ and $j$ are connected, else $0$. The structure of a network in general is dependent upon the way the nodes are connected in a network. However, bus networks fall into a special class of complex networks where the physical constraints offered by the roads and the cities result in the emergence of the network topology~\cite{wang1, wang2, bullock2010spatial, evans2010complex}. The statistical analysis of these networks reveal a wide spectrum of topological structure. We define the degree, $k_i$ of a node, $n_i\in N$ as $\Sigma_i a_{ij}$. The pattern of the inter-nodal connectivity is given by the degree-distribution function, $P(k)$, which can be defined as the probability of a node, $k$ having a degree of atleast $k$. Some of the other network metrics that are of particular interest are the clustering coefficient, $C_{av}$ which denotes the tendency of nodes to form clusters or cliques and the characteristic path length $l_{ij}$. The local clustering coefficient is given by $C (i) =\frac{2|a_{ij}: (n_i , n_j)\in N, a_{ij}\in A_{ij}|}{k_i (k_i - 1)}$, where $a_{ij}$ is the link connecting node pair $(i, j)$, and $k_i$ are the neighbours of the node $n_i$. The neighbourhood for a node, $n_i$ is defined as the set of its immediately connected neighbours, as $\mathcal{N} = \{n_j : l_i \in L \vee l_j \in L\}$, and the average clustering coefficient for the entire network is given by $C_{av}=\Sigma_i C_i /n$. The characteristic path length, $l_{ij}$ is defined as the average number of nodes crossed along the shortest paths for all possible pairs of network nodes. The average distance from a certain vertex to every other vertex is given by $d_i = \Sigma_{i\neq j}\frac{d_{ij}}{|N(G)|-1}$. Then, $l_{ij}$ is calculated by taking the median of all the calculated $d_i$ $\forall i \in \mathcal{R}^n$. Finally, an important network metric is the degree-assortativity that tells us whether the hubs are directly connected in a network or if they are connected through intermediate nodes. We tabulate the statistical properties for the six networks in Table 1. 

We simulate the two epidemic models SI and SIR on these networks. SI model helps us to understand diffusion and percolation in these networks whereas, SIR is the classical model that describes epidemic spreading. Both the models are simulated using agent based modelling technique (igraph), where each node is considered to be an agent whose states (S, I or R) change with every increment in simulation.  

\subsection{SI model}
The SI model is the most basic representation of an epidemic spreading model. In this model, there are two states that an agent or a node can exist in: S (susceptible) or I (infected). The SI model describes the status of individuals or agents switching from susceptible to infected at every instant of time. It is assumed that the population is homogeneous and closed, \emph{i.e.}, no new entity is either created due to birth or removed due to death, and also no new entity enters the system, thus preserving homogeneous mixing in the system. The SI model also implies that each individual has the same probability to transfer disease, innovation or information to its neighbors. Thus, the SI model helps to capture the diffusion or percolation process in the entire network. The SI model is formulated using the following differential equation. Since an agent in the entire population can either be in state S or I, 
\begin{equation}
S + I = 1
\label{eq:01}
\end{equation}
The SI model is governed by a single parameter, $\beta$, the infection transmission rate or simply, the infection rate. The growth in the number of agents in either of the sates is given by:
\begin{equation}
\frac{dS}{dt}=\frac{dI}{dt}=-\beta SI
\label{eq:02}
\end{equation}
Substituting the value of S from equation (1) to equation (2), we get the following differential equation describing the growth rate of I:
\begin{equation}
\frac{dI}{dt}=-\beta (1-I)I
\label{eq:03}
\end{equation}
The solution of the above equation with the initial condition at $t=0$, $I = I_0$ is given by the logistic form:
\begin{equation}
I = (1 + \exp(-\beta t)(\frac{1-I_0}{I_0}))^{-1}
\label{eq:04}
\end{equation}

\subsection{SIR model}
Contrary to the SI model, the agents in SIR model have access to three states S (susceptible), I (infected) and R (recovery). Although the earlier assumptions of a closed population and homogeneous mixing also hold in this case, the complexity of the dynamical process increases due to the addition of one more state. The agents, instead of only switching between susceptible and infected (as in SI model), tend to recover in the SIR epidemic model. The dynamics of the SIR model is controlled by two parameters: the infection rate, $\beta$, and the recovery rate, $\gamma$. The SIR model can be mathematically represented by the set of the following differential equations:
\begin{equation}
S + I + R =1
\label{eqn:05}
\end{equation}
The population of susceptible nodes decrease in proportion to the number of encounters multiplied by the probability that each encounters results in an infection. The negative sign denotes that the population of S is decreasing. Similarly, we can describe the evolution of the other two states, I and R. Nodes become infected at a rate proportional to the number of encounters, and the probability of infection controlled by the parameter, $\beta$. Nodes recover at a rate proportional to the number of infected individuals, and the probability of recovery controlled by the parameter, $\gamma$:
\begin{equation}
\frac{dS}{dt}=-\beta SI,\quad \frac{dI}{dt}=\beta SI - \gamma I\quad and\quad \frac{dR}{dt}=\gamma I
\label{eqn:06}
\end{equation}
It would be interesting to analyze the spread of infection with respect to the susceptible individuals when there is a constant recovery (from equation (6)). We calculate the variation in I with respect to S:
\begin{equation}
\frac{dI}{dS}=\frac{\gamma}{\beta}\frac{1}{S}-1
\label{eqn:07}
\end{equation}
The solution to the above equation with the initial conditions at $t=0$, $I \sim 0$ (negligible as compared to the population) and $S = 1$, is given by:
\begin{equation}
I=\int(\frac{\gamma}{\beta}\frac{1}{S}-1)dS = \frac{\gamma}{\beta}ln(S) - s +1
\label{eqn:08}
\end{equation}
In order to understand the rate of spread of infection in the population, we look at the rate equation for I from equation (6):
\begin{equation}
\frac{dI}{dt} = \beta SI - \gamma I = I(\beta S - \gamma)
\label{eqn:09}
\end{equation}
The above equation implies that the infection spreads if and only if $(\beta S - \gamma) > 0$. The epidemic dies out (the number of infected individuals decreases) if the above quantity is less than zero. Bifurcation occurs at the stationary state, when $\frac{dI}{dt} = 0$, which separates the above two regimes and corresponds to the epidemic threshold. 

\section{Results}

\begin{table}[t]
\begin{tabular}{|cccccccc|}
\hline
{\bf Bus routes} & {\bf Nodes} & {\bf Edges} & {\bf $l_{ij}$} & {\bf $C_{av}$} & {\bf $\gamma$} & {\bf Assortativity} & {\bf $\bar{k}$} \\ \hline
{\bf ABN}        & 1103        & 2582        & 5.59           & 0.19                         & 2.47                      & 0.07              & 3.67           \\
{\bf CBN}        & 1644        & 2732        & 9.02           & 0.142                         & 3.05                       & 0.09              & 3.31          \\
{\bf DBN}        & 1557        & 4287        & 5.51           & 0.18                         & 3.13                 & 0.07              & 9.88           \\
{\bf HBN}        & 1088        & 2954        & 3.87           & 0.26                         & 3.52                      & -0.03             & 23.88          \\
{\bf KBN}        & 518         & 884         & 5.72           & 0.08                         & 4.96                     & -0.01             & 6.72           \\
{\bf MBN}        & 3131        & 6443        & 10.02          & 0.18                         & 3.25                 & 0.45              & 33.38          \\ \hline
\end{tabular}
\caption{Tabular representation of the statistical data for the bus routes of six major Indian cities ($l_{ij}$ = characteristic path length, $C_{av}$ = average weighted clustering coefficient, $\gamma$ = power-law exponent, and $\bar{k}$ = average node degree).}
\label{table}
\end{table}

Both SI and SIR models, although capable of being solved analytically, their exact solutions on a network topology become highly complicated to evaluate due to the stochasticity associated with initial node selection and infection transmission. In order to understand the effect of both the models on a complex network, we need to resort to numerical simulations. We discussed earlier how the structure of the networks are dependent upon the degree-distribution function, $P(k)$. In Table 1, we present the various statistical properties of the networks studied~\cite{chatterjee, chatterjee2015studies}. We observe that the networks follow scale-free degree-distribution patterns with varying power-law exponents, $\gamma$. In Figure 1, we simulate the diffusion dynamics in the networks. The plots show the Cumulative Distribution Function (CDF) of infection transmission in the network (Y-axis) with respect to simulation time (X-axis). As we saw in equation (4), the analytical solution for the SI model gives a logistic curve. However, the rate of diffusion or the slope of the curve and the saturation thresholds will be different for different networks due to their underlying topology. It is interesting to see how the various network metrics affect information diffusion in the following networks. We observe that the characteristic path-length $l_{ij}$ has a direct effect on the diffusion rate in these networks. The above observation is quite obvious as the metric $l_{ij}$ tells us the number of hops that are required to navigate the entire network. From Figure 1, we can observe the simulation time for MBN and HBN by looking at the steepness of the plots. While HBN exhibits the steepest ascent, MBN takes the longest simulation time, which directly correlates to the magnitudes of the characteristic path lengths of MBN and HBN from Table~\ref{table1}

\begin{figure}[t]
\center
\includegraphics[width=1\textwidth]{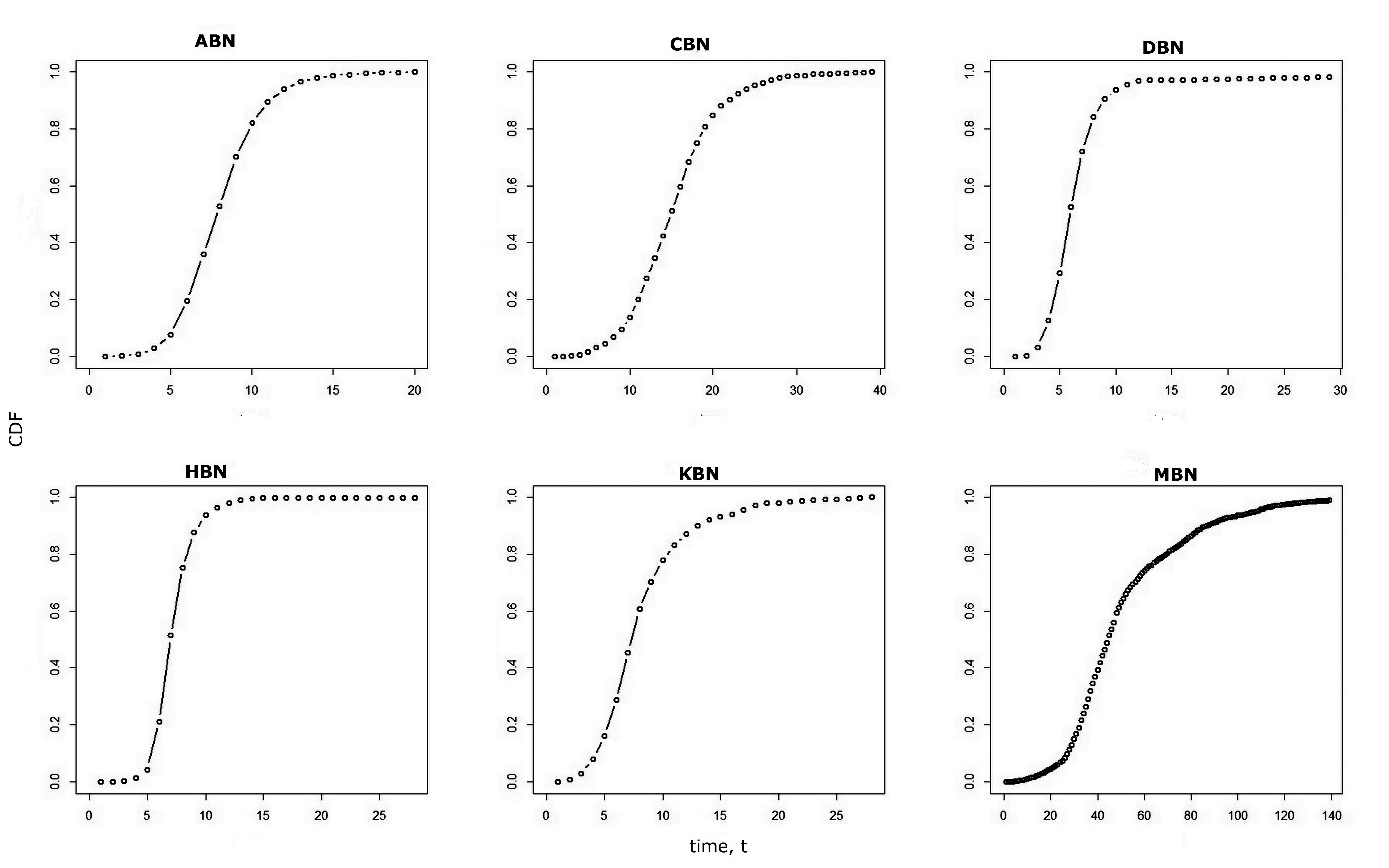}
\caption{SI model simulation on six different networks with $\beta = 0.4$. The Y-axis denotes the Cumulative Distribution Function (CDF) of the infection probability of the nodes, and the X-axis represents simulation time.}
\label{fig:01}
\end{figure}

\begin{figure}[t]
\center
\includegraphics[width=1\textwidth]{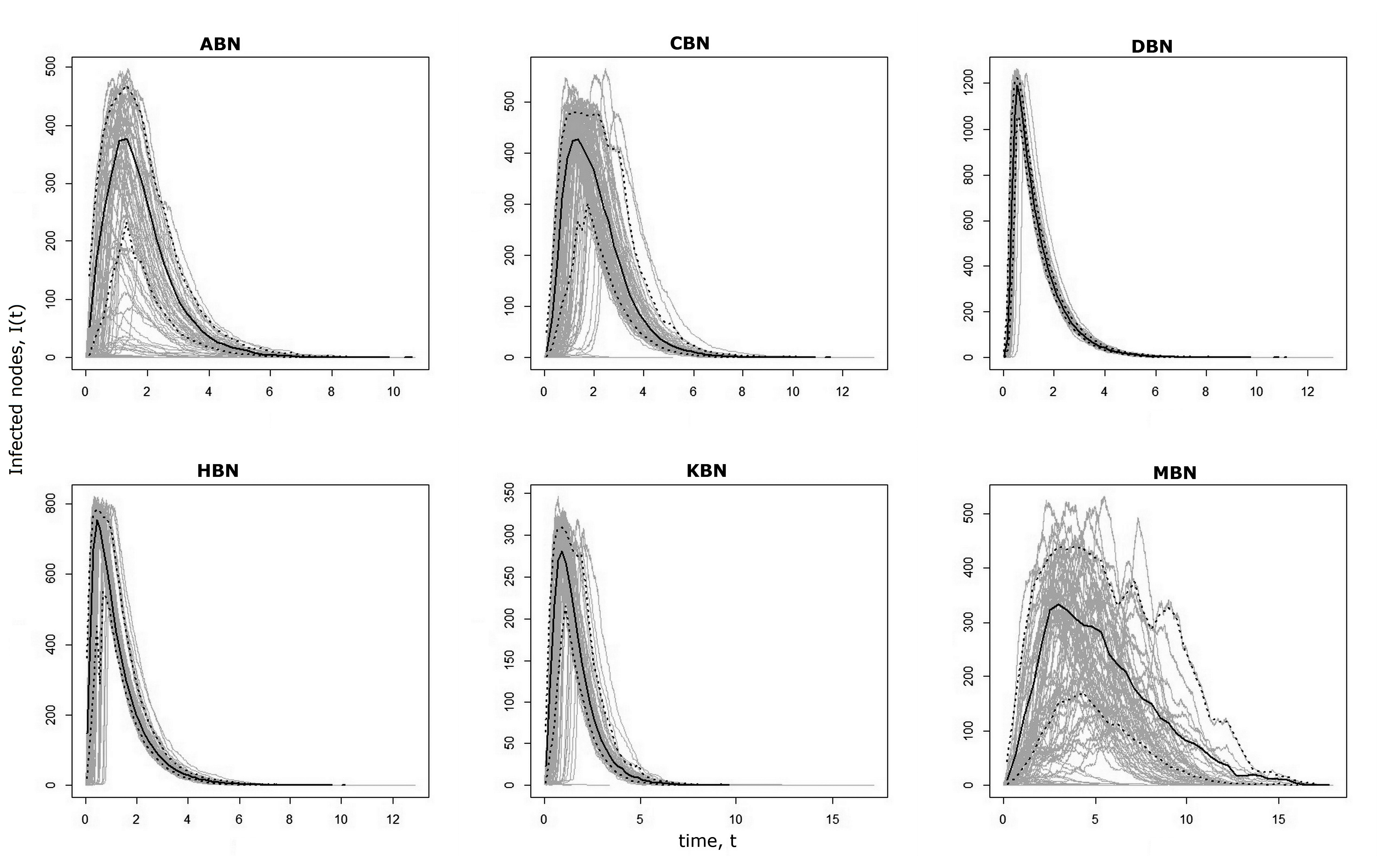}
\caption{SIR model simulation on the six different networks with the parameters, $\beta = 5$ and $\gamma = 1$. The Y-axis denotes the infected nodes and the X-axis represents simulation time. The curves were plotted after 100 simulations; the dark line represents the median distribution, with the dotted ones above and below as the maximum and minimum thresholds.}
\label{fig:02}
\end{figure}

In Figure 2, we plot the SIR simulation results for the different networks. The SIR curve has a typical profile because of the simultaneous decay of the infected individuals and the growth of the recovered individuals. The curve achieves a peak when the recovery rate equals the infection rate. We can clearly observe that the networks which display strong assortative behaviour (CBN and MBN) tend to have multiple peaks. The reason for the presence of multiple peaks can be explained by the fact that assortative networks tend to be hub-attractive, thus infection has multiple pathways to spread across the network, either from hub to hub, hub to node, node to hub or node to node. For weakly assortative and disassortative networks, only three among the above four possibilities exist (excluding hub to hub transmission). An infection transmitting from one hub to another hub is more likely to infect a larger number of nodes than an infection transmitted from a hub to a node. Thus, the threshold values would be achieved very early and the presence of stochasticity in the selection of the initially infected node will induce noise in the plots followed by multiple peaks. Similar to our previous observation, the characteristic path length $l_{ij}$ plays a vital role in the SIR model as well. It can be clearly observed by looking at the steepness of the plots for HBN. While ABN and DBN show similar properties, the peak and the steepness for DBN is much greater than ABN. This can be attributed to the fact that the average degree, $\bar{k}$ in DBN is roughly three times the average degree of ABN (see Table 1). This automatically accelerates the infection transmission rate in DBN, as each node in DBN has three times the number of choices available compared to each node in ABN. 

\begin{figure}[t]
\center
\includegraphics[width=1\textwidth]{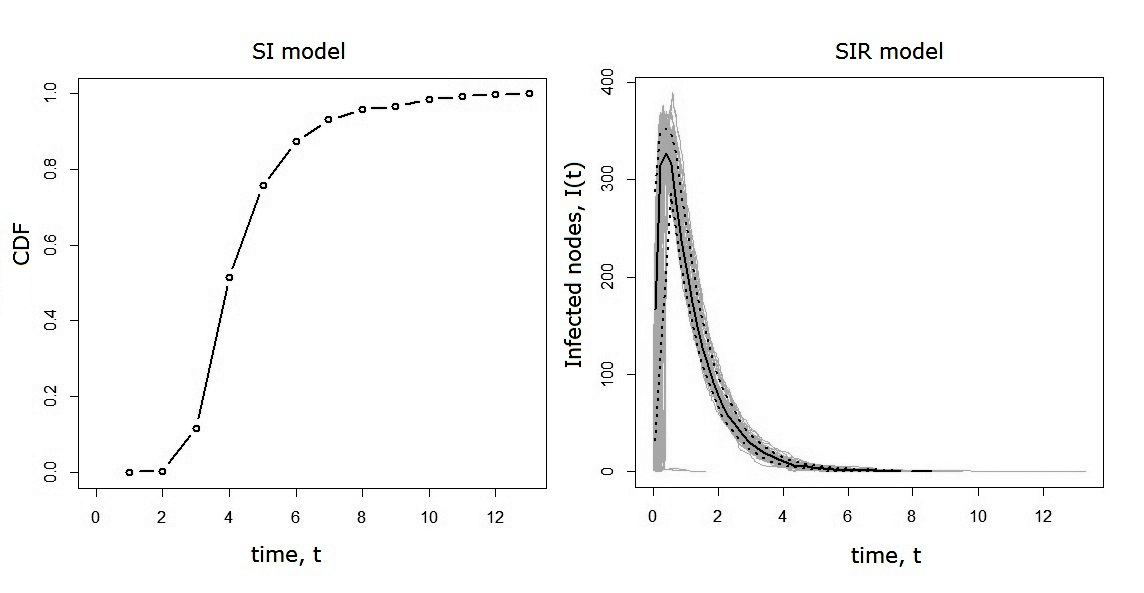}
\caption{SI and SIR simulations on US airline network for 500 of the busiest airports \cite{colizza2007reaction}.}
\label{fig:03}
\end{figure}

\begin{table}[hb]
\centering
\begin{tabular}{|ccccc|}
\hline
             & {\bf $\delta_0$ (s)} & {\bf $\epsilon_0$} & {\bf Sim. time (s)} & {\bf $l_{ij}$} \\ \hline
{\bf ABN}    & 10                              & 0.35                     & 1.6                 & 5.59           \\
{\bf CBN}    & 20                              & 0.42                     & 2                   & 9.02           \\
{\bf DBN}    & 8                               & 0.8                      & 1                   & 5.51           \\
{\bf HBN}    & 8                               & 0.72                     & 1                   & 3.87           \\
{\bf KBN}    & 12                              & 0.55                     & 2                   & 5.72           \\
{\bf MBN}    & 60                              & 0.15                     & 3                   & 10.02          \\
{\bf US air} & 4.5                             & 0.6                      & 1.5                 & 2.92           \\ \hline
\end{tabular}
\caption[Network thresholds for the various networks]{The table outlines the $80\%$ percolation thresholds ($\delta_0$), epidemic thresholds ($\epsilon_0$), their corresponding simulation times and the characteristic path lengths for the various networks.}
\label{table3}
\end{table}

In Figure 3 we simulate the SI and SIR models on the US airline network for 500 of the busiest airports ($N=500, L=2980$)~\cite{colizza2007reaction}. Statistical analysis of the network reveals scale-free degree-distribution pattern between the nodes, with the characteristic path length $l_{ij}=2.92$ and average degree $\bar{k} = 12$. The SI plot of the US airline network and ABN show similar pattern of growth as both of them exhibit scale-free behaviour. However, the SIR plot is similar to that of HBN due to extremely low characteristic path length and high average node degree. In Table 2, we present our findings for the networks studied in the paper. The first column represents the simulation time (in seconds) for $80\%$ percolation threshold from the SI model. In the second and third column, we present the epidemic thresholds for the various networks studied by computing the values of the plots from the SIR model (Figure 2) (as a fraction of network size) and the corresponding simulation times (in seconds) respectively. In the final column, we present the characteristic path lengths for the various networks.

\begin{sidewaysfigure}
\center
\includegraphics[width=1\textwidth]{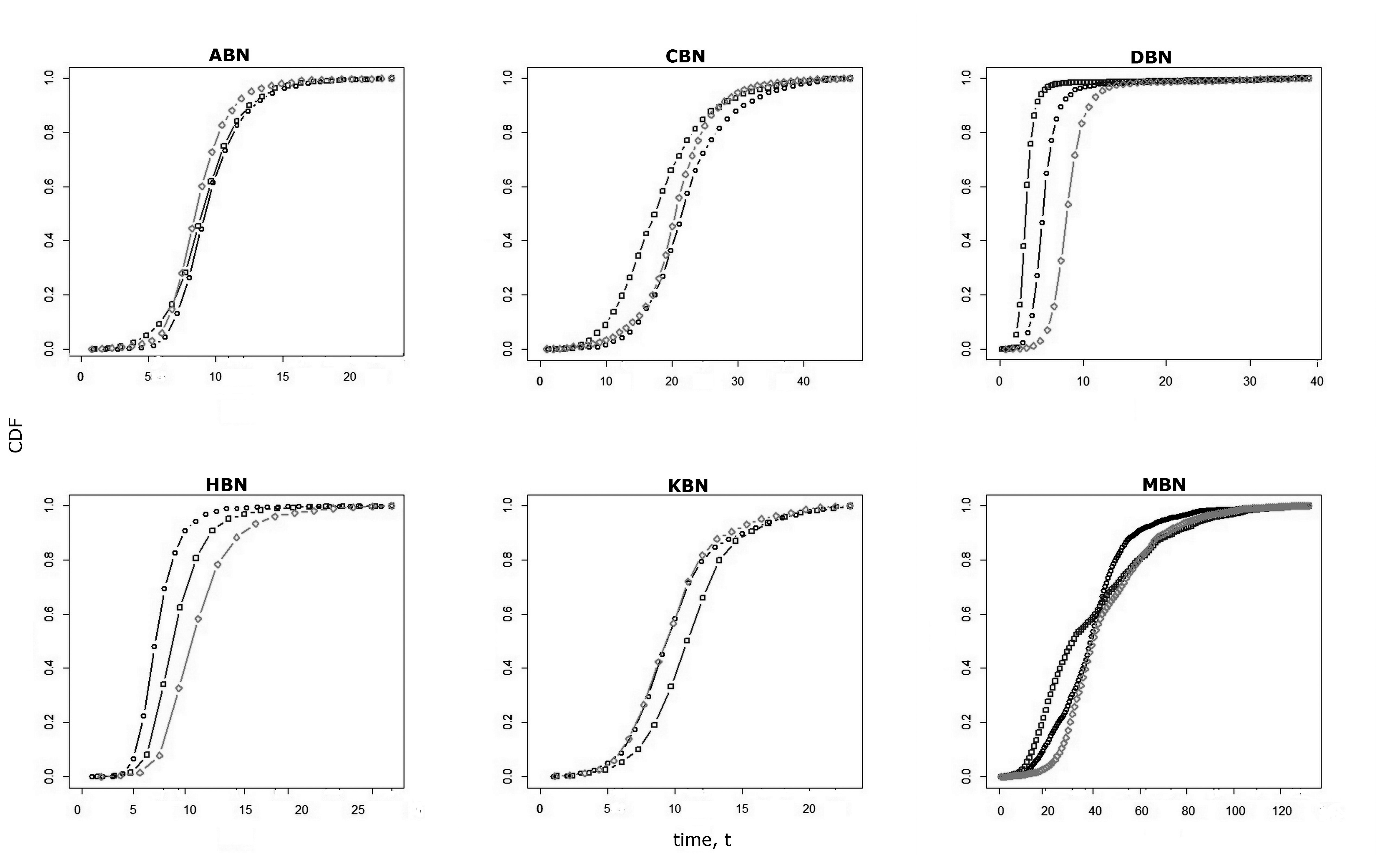}
\caption{SI model simulation on the six different networks after $2\%$ node removal ($\circ$ -- degree-biased, $\Box$ -- betweenness-biased and $\Diamond$ -- closeness-biased). The Y-axis denotes the CDF of the infection probability of the nodes, and the X-axis represents simulation time.}
\label{fig:04}
\end{sidewaysfigure}

Finally, in Figure 4, we plot the variation in the rate of percolation by removing nodes from the network based upon their centralities and degrees. In transportation networks, other than the degree of a node, closeness and betweenness centralities play a crucial role. While closeness centrality is a measure of a node's relative importance in the network due to the existence of shortest paths from that particular node to every other node in the entire network, betweenness centrality, on the other hand, acts as a bridging node connecting different parts of the network together. In order to capture their effects on information diffusion we simulate the SI model on modified networks generated after directed removal of nodes. Since CBN and MBN are strongly assortative, we remove only two percent of the nodes (higher number of node removal will cause CBN and MBN to disintegrate into disconnected components). We find that the removal of nodes does not significantly affect the diffusion in ABN. However for CBN, DBN and HBN, we observe that when nodes are removed based upon their closeness centrality, the diffusion curve shifts towards right, thus signifying a delay in the diffusion process. This can be explained due to the fact that the removal of nodes based upon closeness centrality has a direct effect on the characteristic path length. A node with high closeness allows every other node in the network to be reached along the shortest paths. The removal of such a node affects/delays diffusion until the next central node is encountered. For MBN, we observe that degree-biased removal causes the diffusion rate to increase steeply, signifying the presence of redundant nodes that simply increase the characteristic path length of the network. A removal of $2\%$ of such nodes causes the diffusion to improve significantly, as can be compared from the simulation times recorded in Figure 1 and 4. 

\section{Discussion}
In this paper, we simulated diffusion and epidemic spreading on the bus transportation networks for six different Indian cities, and also compared the results with the US airline network. From Table 2, we can clearly identify the characteristic path length to be a vital component in both, network diffusion and epidemic spreading. Interestingly, the metric $l_{ij}$ can be used as a single parameter to compare diffusion rates across different network topologies. This is because ABN, DBN, HBN and KBN exhibit small-world phenomenon due to which the characteristic path length scales as the logarithm of the network size, \emph{i.e.}, $l_{ij}\sim \log(N)$. However, complexity arises when we have networks with comparable characteristic path lengths, like in the case of ABN, DBN and KBN. In such cases, average node degree plays an important role. Actually, a node having a degree $k$ has $k$ opportunities for infection. For a network with average node degree $\bar{k}$, the rate of change of the susceptible population is given by:
\begin{equation}
\frac{dS}{dt} = -\beta\bar{k}SI
\label{eqn:10}
\end{equation}
The rate of change of the infected and simultaneously recovered individuals is similarly given by:
\begin{equation}
\frac{dI}{dt} = \beta\bar{k}SI - \gamma I \quad and \quad\frac{dR}{dt} = \gamma I
\label{eqn:11}
\end{equation}
Note that the rate of change of the recovered individuals remains the same as before. Substituting the value of I from equation (11) into equation (10), and solving for S gives us the following expression for S in terms of R,
\begin{equation}
S = -\int(\beta\bar{k}SI) dt = \exp(-\beta\bar{k}R)
\label{eqn:12}
\end{equation}
At time $t\rightarrow\infty$, $I \rightarrow 0$ and $S + R \rightarrow 1$. Population of recovered individuals, $R_{\infty}$ is given as, $R_{\infty} = 1 - \exp(-\beta \bar{k} R_{\infty})$. Recovery of the individuals occur in the network if and only if the slope of $R_{\infty}\geq 1$ or $\beta \geq \bar{k}^{-1}$. Similarly, we can observe from Table 2 that the values of epidemic thresholds (DBN, HBN and US air) are also strongly correlated with the characteristic path length. The epidemic threshold for DBN is particularly high compared to its comparable counterparts, ABN and KBN. The reason for this behaviour lies in the fact that DBN has a high average node-degree when compared to ABN, and is assortative when compared to KBN. This allows DBN a high degree of freedom for infection transmission. Finally, the low epidemic threshold values for CBN and MBN can be directly attributed to their considerably higher magnitudes of characteristic path length. Even though CBN and MBN are strongly assortative, the structural advantage of assortativity in diffusion is ruled out due to the presence of long routes mostly comprised of intermediate nodes. The effect of node centralities in information diffusion is also studied in this paper. It is found that networks are sensitive in percolation and diffusion to that particular metric which has a direct effect on the characteristic path length. For some networks, closeness plays a crucial role (CBN, DBN and HBN). However for networks like MBN, degree-biased removal also reduces the magnitude of $l_{ij}$. Although, the specific node centrality directly affecting the magnitude of $l_{ij}$ largely depends upon the network structure, it can be argued that closeness of a node can be used as a marker for network immunization procedure. 

\section{Conclusion}
We studied the functionality of the bus networks of six major Indian cities in this paper. Since experiments with epidemic outbreaks in a population (or a network) is not a viable option, we resort to mathematical modelling. We, therefore, study the effect of percolation and epidemic spreading on these networks using SI and SIR epidemic models through numerical simulations. While it is observed that the characteristic path length plays a crucial role in information diffusion and epidemic spreading, several other network metrics also play important roles. Their importance is however restricted to their relative contribution to the topological structures of the networks. Small-world property, while an extremely desirable property in transportation networks, is highly subjective in its role of information diffusion, solely due to the diffusing entity. While diffusion of pathogens is an undesirable phenomenon, diffusion of useful information is a desirable one. In conclusion, the work presented in this paper will help us in understanding and controlling the process of diffusion and epidemic outbreaks in spatially-constrained networks. 

\section*{Acknowledgments}
The authors acknowledge the support from the Information Technology Research Academy, a Division of Media Labs Asia, a non-profit organization of the Department of Electronics and Information Technology, funded by the Ministry of Communications and Information Technology, the Government of India and the Center of Excellence in Urban Transport at the Indian Institute of Technology, Madras, sponsored by the Ministry of Urban Development, the Government of India.

\end{document}